\documentclass[prl,twocolumn,floatfix,showpacs,superscriptaddress]{revtex4}

\usepackage{amsmath,amsfonts,amssymb,bm}
\usepackage{dcolumn}
\usepackage[final]{graphicx}
\usepackage{bm}
\usepackage{comment}

\includecomment{pdffigure}

\begin{document}

\bibliographystyle{apsrev}	

\title{Gap plasmonics of silver nanocube dimers}

\author{Dario Knebl}
\author{Anton H\"orl}
\author{Andreas Tr\"ugler}

\affiliation{Institute of Physics,
  University of Graz, Universit\"atsplatz 5, 8010 Graz, Austria}

\author{Johannes Kern}

\affiliation{Institute of Physics and Center for Nanotechnology,
   University of M\"unster, 48149 M\"unster, Germany}

\author{Joachim R. Krenn}
\author{Peter Puschnig}
\author{Ulrich Hohenester}
\email{ulrich.hohenester@uni-graz.at}

\affiliation{Institute of Physics,
  University of Graz, Universit\"atsplatz 5, 8010 Graz, Austria}

\date{\today}

\begin{abstract}
We theoretically investigate gap plasmons for two silver nanocubes coupled through a molecular tunnel junction.  In absence of tunneling, the red-shift of the bonding mode saturates with decreasing gap distance.  Tunneling at small gap distances leads to a damping and slight blue-shift of the bonding mode, but no low-energy charge transfer plasmon mode appears in the spectra.  This finding is in stark contrast to recent work of Tan et al. [Science \textbf{343}, 1496 (2014)].
\end{abstract}



\maketitle


Gap plasmonics \cite{esteban:12} deals with surface plasmons (SPs) \cite{maier:07} in narrow gap regions of coupled metallic nanoparticles.  For sufficiently narrow gaps, electrons can tunnel directly from one nanoparticle to the other one, leading to the emergence of new charge transfer plasmons (CTPs) \cite{esteban:12,savage:12,duan:12,scholl:13,esteban:15}.  Molecular tunnel junctions enable tunneling over larger gap distances in the nanometer regime~\cite{tan:14,benz:15}, and thus establish a novel platform for hybrid structures reconciling molecular electronics with plasmonics.

Recent years have seen significant research efforts to understand the properties of gap plasmons, and have highlighted the importance of the tunneling strength as a trigger for the CTP appearance~\cite{wu:13} and of the gap morphology which strongly influences the CTP spectral position~\cite{esteban:15b}: for rounded gap terminations the bonding mode red-shifts with decreasing gap separation, until tunneling sets in, as evidenced by the appearance of a low-frequency CTP together with a broadening and blue-shift of the bonding mode \cite{esteban:12,savage:12,esteban:15}.  In contrast, for flat terminations the red-shift of the bonding mode saturates with decreasing gap distance, while at the same time the transversal cavity plasmon (TCP) modes shift to the red; here, the onset of tunneling has no significant impact on the bonding mode and no low-frequency CTP appears in the spectra.  

In this paper, we theoretically investigate the plasmonic properties of two coupled silver nanocubes, similarly to the electron energy loss spectroscopy (EELS) experiments of Tan et al.~\cite{tan:14} for two nanocubes coupled through a molecular tunnel junction.  We compute EEL and extinction spectra using the MNPBEM toolbox~\cite{hohenester.cpc:12,hohenester.cpc:14b,waxenegger:15}, supplemented with the quantum corrected model (QCM)~\cite{hohenester.prb:15} to account for quantum tunneling.  We find that the red-shift of the bonding mode saturates with decreasing distance and an additional tunnel conductivity in the gap region leaves the spectral position unaffected.  The TCP modes shift with decreasing gap distance to the red, and the tunnel conductivity damps these modes.  All these findings are in perfect agreement with the observations of Esteban et al.~\cite{esteban:15} for flat gap terminations and would qualify our work as a systematic research paper, if it was not for this single point: despite serious efforts we were unable to confirm the emergence of the low-energy CTP observed by Tan et al.~\cite{tan:14} and could not reproduce their simulation results.  We will argue why we believe that our results are valid within the electrodynamic and QCM model under consideration, and why a re-interpretation of the experiments might be needed.


In our simulations we model the cubes with rounded edges and corners as superellipsoides, whose boundaries are parameterized through $u\in[0,\pi)$ and $v\in[-\pi,\pi)$ according to
\begin{subequations}\label{eq:superellipsoid}
\begin{eqnarray}
  x(u,v)&=&a\,s(u,r)\,c(v,r)\\
  y(u,v)&=&a\,s(u,r)\,s(v,r)\\
  z(u,v)&=&a\,c(u,r)\,,
\end{eqnarray}
\end{subequations}
where $a$ determines the cube size (we use side lengths of 35 nm throughout), $r$ is a rounding parameter, and we have introduced the functions $s(\xi,r)=\mbox{sign}(\sin\xi)|\sin\xi|^r$ and $c(\xi,r)=\mbox{sign}(\cos\xi)|\cos\xi|^r$.  For the cubes we set $r=0.25$, but will later use larger $r$ values in order to morph the cubes to spheres \cite{schmidt:14b}.  For the electrodynamic simulations we employ the MNPBEM toolbox~\cite{hohenester.cpc:12,hohenester.cpc:14b,waxenegger:15} and use for the dielectric function of silver tabulated values extracted from optical experiments~\cite{johnson:72}.

\begin{figure}
\begin{pdffigure}
\centerline{\includegraphics[width=\columnwidth]{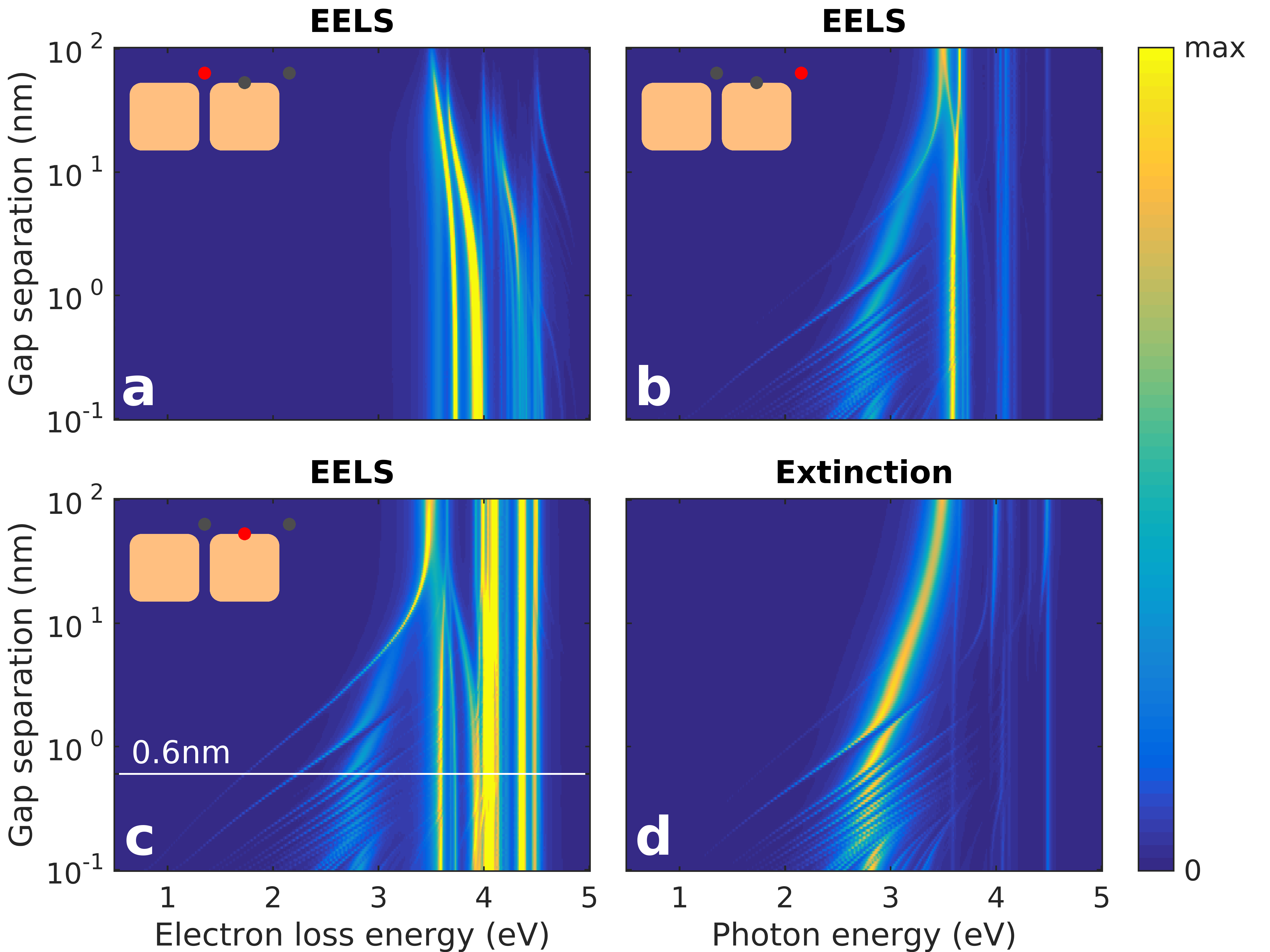}}
\end{pdffigure}
\caption{(Color online) Classical electrodynamic simulations of (a--c) EEL and (d) optical extinction spectra for two coupled silver nanocubes with side lengths of 35 nm.  The impact parameters for the electron beams are indicated in the insets of panels (a--c), and the light polarization direction is along the cube connection direction in panel (d).  Notice the logarithmic scale used for the gap distances.  All density plots are normalized to the respective maximal values.  The line in panel (c) reports the gap distance used in Fig.~2.}
\end{figure}

Figure 1 shows density plots of the (a--c) EEL and (d) optical extinction spectra for two coupled silver nanocubes as a function of gap distance, using classical electrodynamic simulations where no tunneling is considered.  For the EEL spectra the impact parameters of the electron beams are indicated in the insets, and for the optical spectra the light polarization is along the direction of the cube connection.  For large gap separations the EEL and optical spectra agree with those of a single cube, whose modes have been studied in detail elsewhere~\cite{sherry:05,nicoletti:13}.  With decreasing gap distance the bonding mode (denoted in Ref.~\cite{esteban:15b} also as longitudinal antenna plasmon, LAP) shifts to the red~\cite{bordley:15}, as seen most clearly in the extinction spectra of Fig.~1(d).  At distances around a few nanometers new modes appear in the spectra which continuously red-shift when further decreasing the gap distance.  In accordance to Ref.~\cite{esteban:15b}, and as shown by the surface charge maps in Fig.~2(c), we assign these modes to TCPs.  Whenever these modes cross the bonding mode we observe a clear anti-crossing, a finding which we attribute to mode coupling.  The overall red-shift of the bonding mode saturates for the smallest gap distances, say at a value of 2.8~eV.  From the comparison of the different panels of Fig.~1 we see that these mode characteristics can be observed in both EEL and optical spectra, with the only exception of panel (a), where the electron beam is located in the center plane of the gap and the excitation of the bonding mode is forbidden because of symmetry~\cite{hohenester.prl:09}.

\begin{figure}
\begin{pdffigure}
\centerline{\includegraphics[width=\columnwidth]{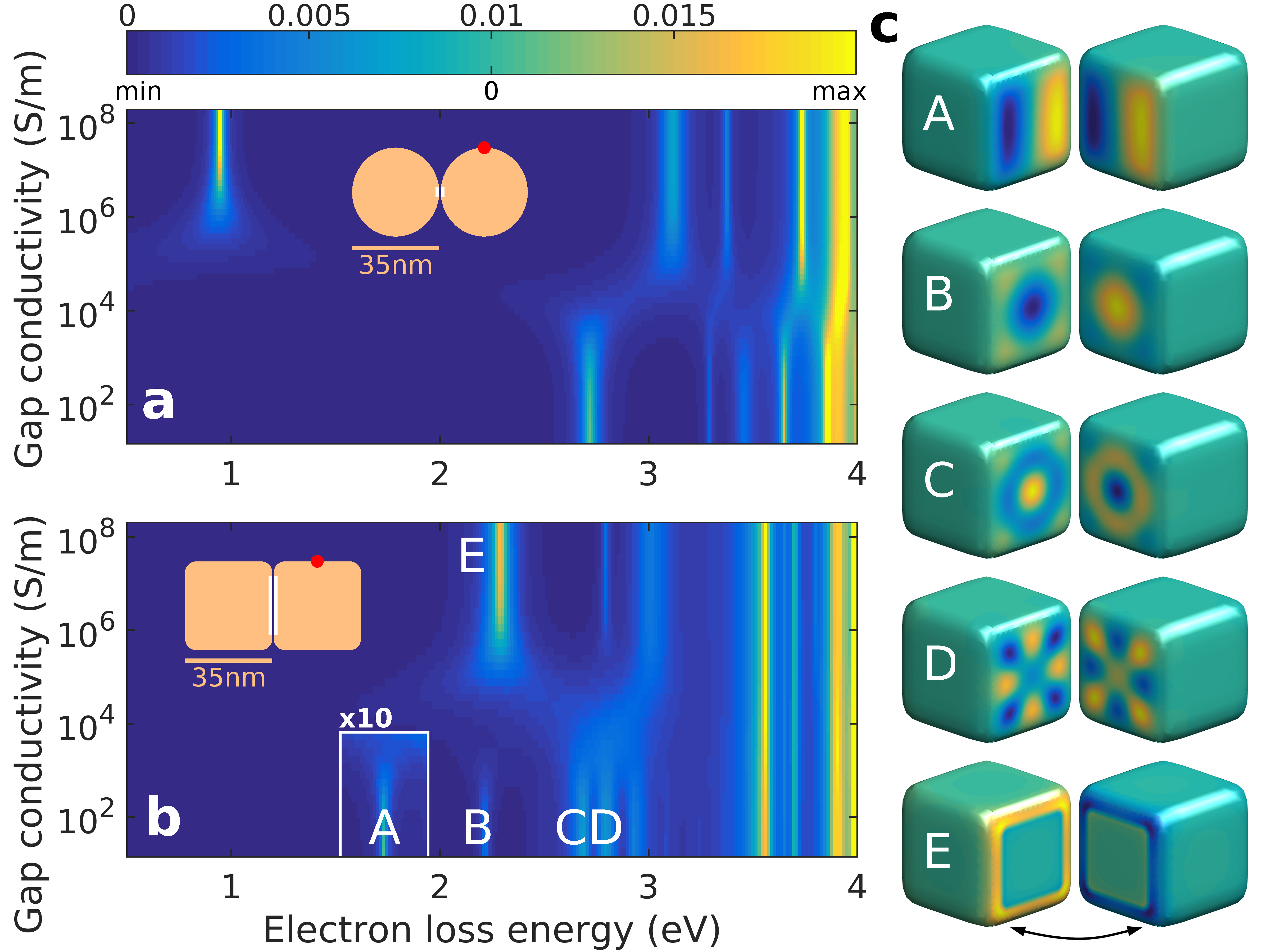}}
\end{pdffigure}
\caption{(Color online) Density plots for EEL spectra of coupled silver (a) spheres and (b) cubes as a function of loss energy and  gap conductivity, and for a gap distance of 0.6 nm.  The impact parameters of the electron beams and the active region where tunneling is considered are shown in the insets.  The color bar indicates the loss probabilities in eV$^{-1}$.  Probabilities in region A have been multiplied by a factor of ten for clarity.  (c) Surface charge distribution (imaginary part) of modes A--E at the resonance energies.  Cubes are rotated apart to offer a better view to the gap region.}
\end{figure}

Figure 2 shows density plots of EEL spectra (electron beam positions indicated in insets) for two coupled (a) spheres and (b) cubes separated by a distance of 0.6 nm.  We allow for tunneling within a distance region of 0.6--0.8 nm (see inset, distance region chosen in order to mimic molecular tunnel junction) using the QCM of Ref.~\cite{hohenester.prb:15}.  In each simulation the gap conductivity within the region where tunneling is allowed is set to a constant value.  For the spheres shown in panel (a) and for the smallest gap conductivities $\sigma_{\rm gap}$, the lowest SP mode at an energy of 2.7 eV is attributed to the bonding mode.  When increasing $\sigma_{\rm gap}$, above a critical threshold of say $10^5$ S/m there is a transition where (i) a CTP appears at an energy of about 1 eV and (ii) the bonding mode blue-shifts and broadens.  These features are in agreement with the literature~\cite{esteban:12,esteban:15,hohenester.prb:15}.  Also the weak dependence of the SP energy on $\sigma_{\rm gap}$ above or below the critical threshold has been previously reported~\cite{wu:13}.

For the coupled nanocubes shown in Fig.~2(b) there is again a transition in the EEL spectra when increasing $\sigma_{\rm gap}$, and again above or below the critical threshold the SP energies depend very weakly on the gap conductivity.  As regarding the SP modes, we observe above the critical $\sigma_{\rm gap}$ value the appearance of a new mode E, which, in contrast to the spheres, is \textit{not} accompanied by an additional low-energy CTP mode.  This finding is in agreement with that of Esteban et al.~\cite{esteban:15b} for flat gap terminations, and highlights the importance of the gap morphology on the SP modes.  

In Fig.~2(c) we report the surface charge distributions of a few selected SP modes.  For small $\sigma_{\rm gap}$ values, A--D correspond to hybridizations between TCP and bonding modes.  In principle, because of symmetry all modes are double or multiple degenerate~\cite{langbein:76} and the mode symmetry shown in the figure is governed by the electron beam position.  Above the critical $\sigma_{\rm gap}$ threshold, (i) the cavity modes become damped (see for instance disappearance of mode A in Fig.~2(b), whose intensity has been magnified by a factor of 10 for clarity), and (ii) a new mode E appears which dominates in the EEL spectra.  As can be inferred from Fig.~2(c), mode E is a CTP where electron tunneling leads to an opposite charging of the cubes.

\begin{figure}
\begin{pdffigure}
\centerline{\includegraphics[width=\columnwidth]{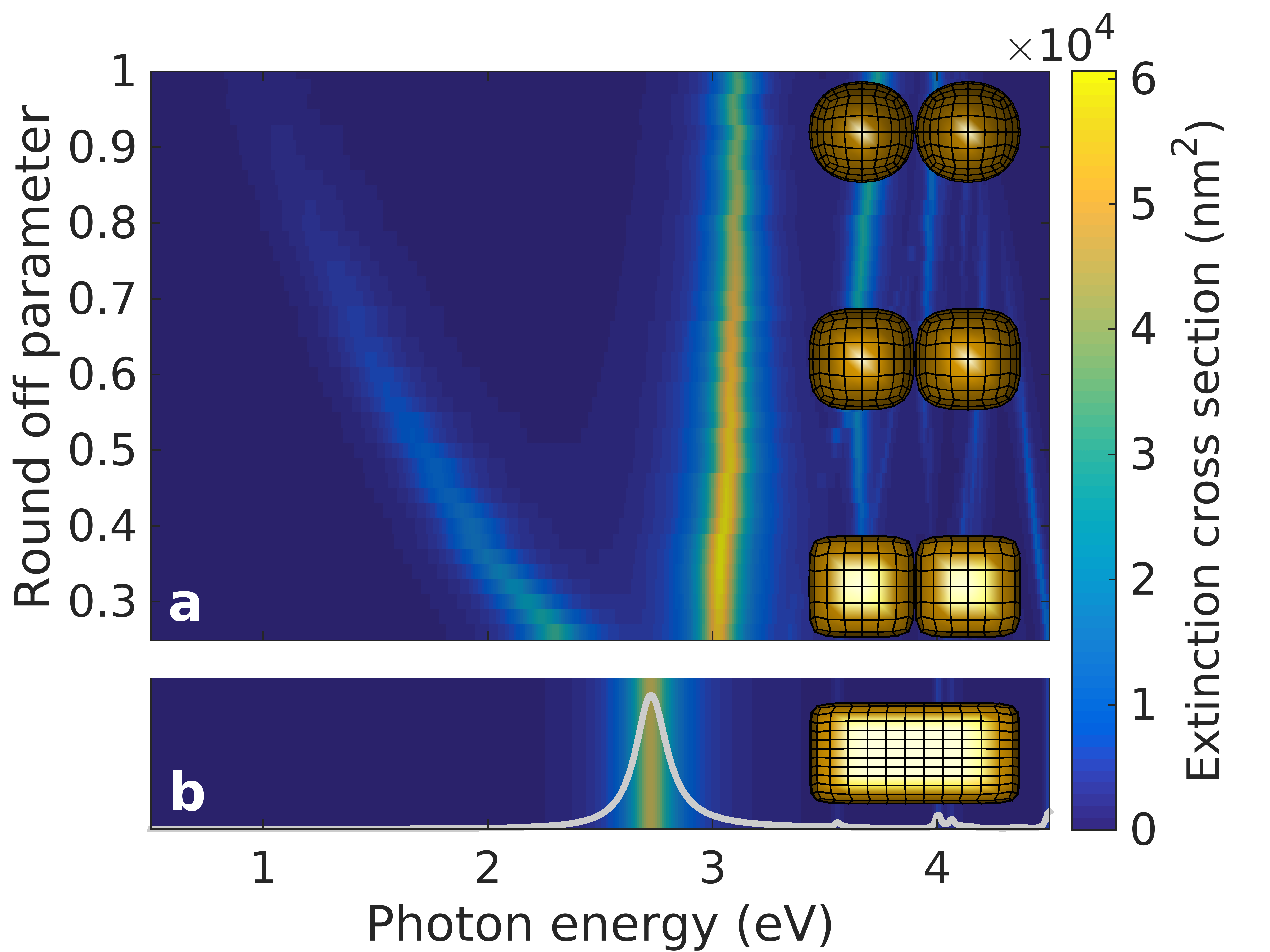}}
\end{pdffigure}
\caption{(Color online) (a) Morphing from two coupled spheres (rounding parameter $r=1$) to two coupled cubes ($r=0.25)$.  The density plot reports the extinction cross section for a gap separation of 0.6 nm, for a light polarization along the nanoparticle connections, and for a tunnel conductivity of $2.49\times 10^5$ S/m representative for BDT.  Similarily to Fig.~2, we consider tunneling within a distance region of 0.6--0.8 nm, as further discussed in the text.  In the insets we report the geometries for three selected structures.  (b) Extinction spectrum for a cuboid whose side length is twice the cube side length.
}
\end{figure}

To further explore the impact of the gap morphology on the SP energies, in Fig.~3 we investigate the scenario where two coupled spheres are deformed to two coupled cubes.  Such morphing has been proven successful for a deeper insight to SP mode characteristics~\cite{schmidt:14b}.  In our simulations we vary the rounding parameter $r$ in Eq.~\eqref{eq:superellipsoid} from 0.25 for the cubes to 1 for the spheres.  The gap distance is set to 0.6 nm for all geometries, and we again consider tunneling within a distance region of 0.6--0.8 nm using a tunnel conductivity of $2.49\times 10^5$ S/m representative for BDT.  For the spheres with $r=1$ we observe in the extinction spectra of Fig.~3(a) the CTP and bonding modes at energies of 1 eV and 3 eV, respectively.  Upon morphing to two cubes, (i) the CTP mode shifts to higher energies and (ii) the bonding mode acquires a higher oscillator strength.  For comparison, in Fig.~3(b) we show the extinction spectrum for a cuboid with a side length of twice the cube length, consisting of one major peak approximately at the energy position of the CTP mode for the coupled nanocubes.  Similarly, it has been shown that the CTP peak for the coupled spheres has approximately the same energy as the dipole mode for two slightly coalescing spheres (``negative gap distance'')~\cite{esteban:12,esteban:15}. 

\begin{figure}
\begin{pdffigure}
\centerline{\includegraphics[width=\columnwidth]{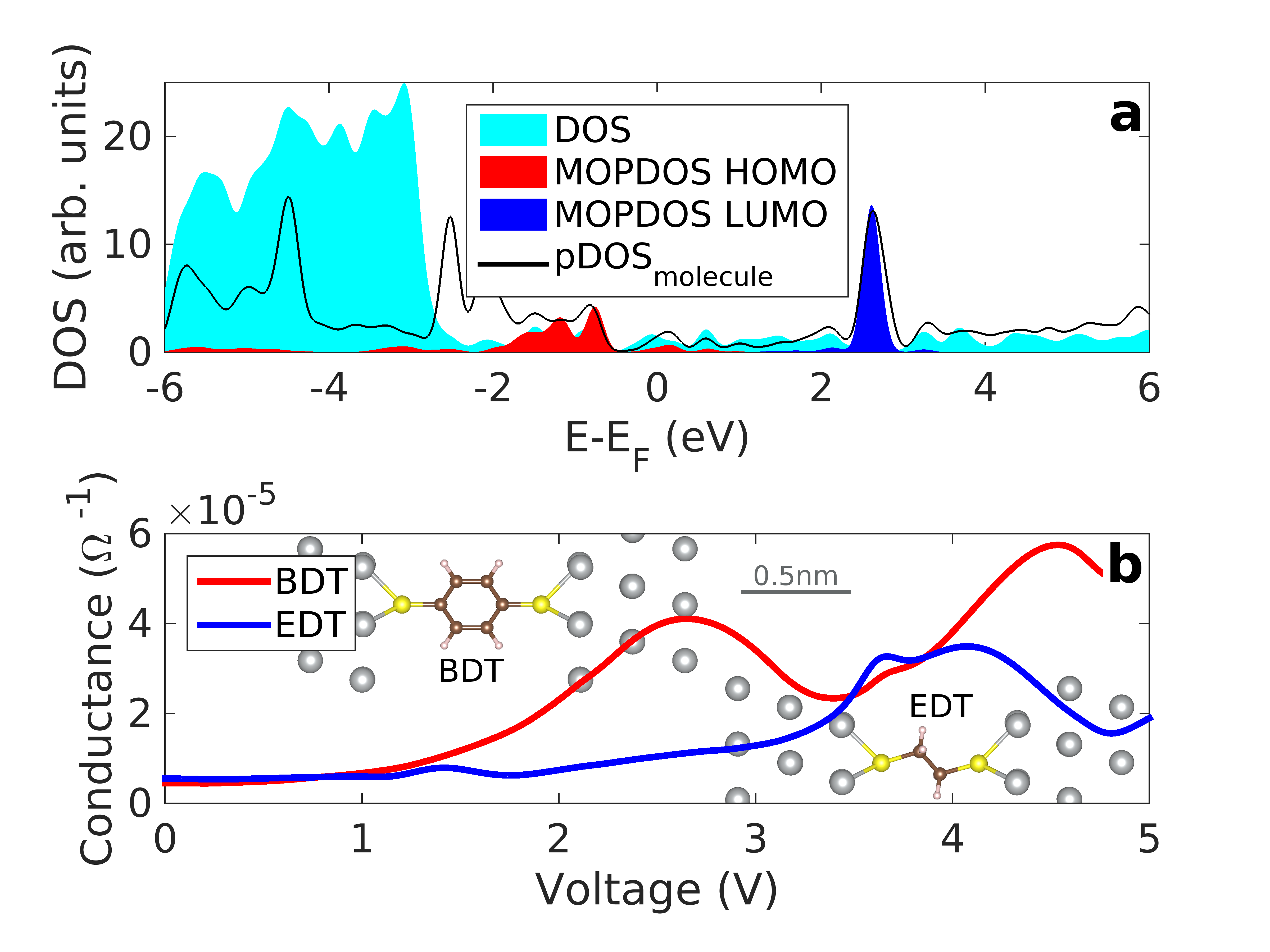}}
\end{pdffigure}
\caption{(Color online) Density functional theory (DFT) simulations for the conductance through the BDT and EDT molecules.  (a) Density of states (DOS) for BDT junction as obtained from the VASP code~\cite{kresse:93,kresse:99}.  We show the total and projected DOS (see text for details).  (b) Conductance through molecular BDT and EDT junctions (see inset for simulated structures) as computed with the TRANSIESTA code~\cite{soler:02,brandbyge:02}.}
\end{figure}

We finally analyze the tunnel conductivities of the molecular junctions of Tan et al.~\cite{tan:14} consisting of aromatic 1,4-benzenedithiolates (BDT) and saturated aliphatic 1,2-ethanedithiolates (EDT) molecules.  The authors have estimated values of $2.49\times 10^5$ S/m for BDT and $9.16\times 10^4$ S/m for EDT.  As a separate estimate for these values, we have calculated the ground state electronic structure and transport properties of the BDT and EDT junctions by ab-initio density functional calculations.  In a first step, we have relaxed the junction geometries and computed the ground state electronic structure by adopting a repeated slab approach using five silver layers on each side of the junction.  For these calculations we have used the VASP code~\cite{kresse:93,kresse:99} employing projector augmented wave (PAW) potentials and have optimized the gap separation, the geometry of the two topmost Ag layers on each side of the junction, as well as all molecular coordinates.  The resulting geometries for both types of molecules and the electronic structure for the BDT junction are depicted in panels (a) and (b) of Fig.~4, respectively.  From the density of states projected onto the molecular orbital of the free molecules (MOP-DOS), we see that the LUMO of BDT, located 2.5 eV above the Fermi level, only weakly hybridizes with the silver surface, while the HOMO is spread between $-2.0$ and $-0.5$ eV below $E_F$ indicating a stronger hybridization with the substrate.  The overall DOS is dominated by Ag $d$-states which appear at a binding energy of about $-3.0$ eV.  In a second step, we have computed the ballistic electron transport through the molecular junctions by using the TRANSIESTA code~\cite{soler:02,brandbyge:02} which is based on the Landauer-B\"uttiker formalism. Using a double zeta, polarized (DZP) basis set, which has been validated by comparing with our VASP DOS results, we have computed the conductance through BDT and EDT junctions as a function of the junction bias, as depicted in panel (b).  At low voltages the conductance of EDT is slightly larger than that of BDT due to the  smaller junction width of the latter.  At bias voltages above 1 and 3 eV, the BDT junction clearly exhibits a larger conductance owing to the fact that the frontier HOMO and LUMO orbitals are located closer to $E_F$ in BDT as compared to EDT.  The low-voltage conductance relevant for the small electric fields of EELS excitations is about $0.5\times 10^{-5}$ S for both BDT and EDT, which corresponds to $0.0645\,G_0$ in units of the conductance quantum $G_0$.  This value is somewhat smaller than the estimated $0.46\,G_0$ (BDT) and $0.20\,G_0$ (EDT) of Ref.~\cite{tan:14}, but is of the same order of magnitude, although one can expect that misalignement of molecules in the junction or finite temperatures will lead to even smaller values~\cite{kim:11}.


We are now in the position to critically examine the work of Tan et al.~\cite{tan:14}.  First, our results are in disagreement with their finite element method (FEM) simulations which showed in the extinction spectra an additional peak at photon energies below 1 eV that was interpreted as a CTP.  In this work we have motivated why such a low-energy peak should not appear in the spectra (we additionally performed finite difference time domain ---FDTD--- simulations with the Lumerical software, for rounded nanocubes with and without a conductivity layer in between the cubes, to confirm the absence of such a peak).  Our most striking argument, in agreement to Esteban et al.~\cite{esteban:15b}, concerns the morphology of the gap:  as can be clearly seen in Fig.~3, the modification of the gap termination from round (spheres) to flat (cubes) comes along with a continuous blue-shift of the CTP, whose energy finally falls together with that of the bonding mode.  Additionally, for dimers with ``negative gap distances'', i.e., coalescing spheres or a cuboid with a side length of twice the cube side length, the SP energies of the dipole modes approximately agree with those of the CTPs.  As the cuboid has a dipole SP energy at about 2.7 eV, we exclude the possibility of a sub-eV CTP for two tunnel-coupled cubes.

After submission of our work we became aware~\cite{nijhuis:15} that Tan et al.~do not use a constant tunnel conductivity $\sigma_0$, but rather a frequency dependent expression $\sigma(\omega)=\sigma_0/(1-i\omega\tau)$ that corresponds to a Drude-type permittivity
\begin{equation}\label{eq:drude}
  \varepsilon_{\rm Drude}(\omega,\ell)=1-\frac{\tilde\omega_p^2}{\omega(\omega+i/\tau)}\,,\quad
  \tilde\omega_p=\sqrt\frac{\sigma_0}{\varepsilon_0\tau}\,,
\end{equation}
with $\tau$ being a collision time.  The effective plasma frequency $\tilde\omega_p$ depends on the conductivity $\sigma_0$ which is computed from quantum mechanical tunneling theory~\cite{wu:13,tan:14}.  Inserting the permittivity of Eq.~\eqref{eq:drude} into our BEM simulations and using a collision time $\tau=30$ fs, representative for silver, we indeed observed a low-energy peak in our EEL and extinction spectra.  It should be noted first that the use of Eq.~\eqref{eq:drude} was previously not mentioned~\cite{tan:14} and that related work for molecular tunnel junctions used a constant $\sigma_0$~\cite{benz:15}, in accordance to our approach.

So why does $\varepsilon_{\rm Drude}$ give a low-energy peak in contrast to a frequency independent $\sigma_0$?  We believe that the low-energy peak in the simulations is due to collective excitations $\mbox{Re}[\varepsilon_{\rm Drude}(\omega\approx\tilde\omega_p)]=0$ built into the Drude model.  These resonances correspond to bulk plasmons of the (fictitious) charge carriers of the tunnel material.  Setting for silver $\hbar\omega_p=9$ eV and $\sigma_{\rm Ag}=6.3\times 10^7$ S/m, we get for the BDT conductivity an effective plasmon energy $\hbar\tilde\omega_p=\hbar\omega_p\sqrt{\sigma_0/\sigma_{\rm Ag}}\approx 0.6$~eV which is similar to the CTP energy found by Tan et al.~\cite{tan:14}.  To make things clear, this resonance has nothing to do with a CTP or any type of plasmonic enhancement, but is a genuine absorption peak of the tunnel material.  Indeed, we found EEL and extinction peaks at precisely the same energy for tunnel-coupled spheres or planar layers.

We next argue why we consider a constant $\sigma_0$ to be a much more reasonable choice.  First, the conductivity in the molecular tunnel junction is due to tunneling and not to free carriers subject to collisions.  In the static case one can compute $\sigma_0$ from tunneling theory~\cite{esteban:12,wu:13,esteban:15} or in the (related) Landauer-B\"uttiker formalism built into the TRANSIESTA code, as we do in our work.  In the time dependent case and for small frequencies, we can adopt the same reasoning as Esteban et al.~\cite{esteban:12,esteban:15} and assume that the modulation of the electric field is slow in comparison to the tunnel process, such that we can describe the system quasi adiabatically (coming back to the static case).  In this approximation, which we assume to be valid in the sub-eV regime, the field is slowly changing and electrons tunnel in presence of the respective field.  This approximation yields a constant $\sigma_0$.  It is also unclear to us why one should describe tunneling using a collision time $\tau$.  How would one interpret these collisions?  And which value should be chosen for $\tau$?  Finally, even if $\sigma_0$ has a frequency dependence, say even by a few orders of magnitude, Fig.~2 shows that this would not change dramatically our conclusions: conductivity only triggers the appearence of the CTP peak, but has otherwise no dramatic impact.  For all these reasons we think that the interpretation of the low-energy peak in the EEL spectra of Tan et al.~in terms of a CTP is not justified by the simulation results, thus calling for a reconsideration of the experimental findings.  We hope that our work will trigger further research in this direction.

\textit{Acknowledgment}.---This work has been supported in part by the Austrian science fund FWF under the SFB F49 NextLite and NAWI Graz.  We gratefully acknowledge Rudolf Bratschitsch for granting access to the FDTD software.


\end{document}